\documentclass{pasj00}
\usepackage{txfonts}
\begin{document}
\SetRunningHead{Author(s) in page-head}{Running Head}
\Received{2010/04/02}%{yyyy/mm/dd}
\Accepted{2010/06/21}%{yyyy/mm/dd}

\title{Redshifts of distant blazar limited by Fermi and VHE $\gamma$-ray observations}

%%% begin:list of authors
% Do NOT capitalize all letters in "textsc".
\author{Jianping \textsc{Yang} %
  \thanks{National Astronomical Observatories, Yunnan Observatory,
Chinese Academy of Sciences,  Kunming 650011, China}}
\affil{National Astronomical Observatories, Yunnan Observatory,
Chinese Academy of Sciences,  Kunming 650011, China; Key Laboratory
for the Structure and Evolution of Celestial Objects, Chinese
Academy of Sciences,  Kunming 650011, China; Yunnan Agricultural
University, Kunming 650201, China} \email{yangjp@ynao.ac.cn}

\author{Jiancheng \textsc{Wang}}
\affil{National Astronomical Observatories, Yunnan Observatory,
Chinese Academy of Sciences,  Kunming 650011, China; Key Laboratory
for the Structure and Evolution of Celestial Objects, Chinese
Academy of Sciences,  Kunming 650011, China}
\email{jcwang@ynao.ac.cn}
%\email{ccccc@xxx.xxx.xx.xx}
%%% end:list of authors

%%% Please use the following style in case that sorting by
%%% affilation is impossible.
%
% \author{%
%   D-Firstname \textsc{D-Familyname}\altaffilmark{1}
%   E-Firstname \textsc{E-Familyname}\altaffilmark{1,2}
%   and
%   F-Firstname \textsc{F-Familyname}\altaffilmark{2}}
% \altaffiltext{1}{Address of Institute}
% \email{ddddd@xxx.xxx.xx.xx}
% \email{eeeee@xxx.xxx.xx.xx}
% \altaffiltext{2}{Address of Institute}

%% `\KeyWords{}' always has to be placed before `\maketitle'.
\KeyWords{gamma-rays:  observations --- BL Lacertae objects:
individual: PG 1553+113, 3C 66A, PKS 1424+240 --- diffuse radiation } %Do NOT move this preamble from here!

\maketitle

\begin{abstract}
Our goal is to limit the redshifts of three blazars PG 1553+113, 3C
66A and PKS 1424+240, through the investigation of their Fermi and
VHE (very high energy) $\gamma$-ray observations. We assume that the
intrinsic spectra of PG 1553+113, 3C 66A, and PKS 1424+240 have not
any cutoff across the Fermi and VHE $\gamma$-ray energy ranges. The
intrinsic spectra of VHE $\gamma$-rays are obtained through the
extrapolation of Fermi spectra. Comparing the measured and intrinsic
VHE spectra due to extragalactic background light (EBL) absorption,
we give the redshift upper limits of three blazars assuming a
specific EBL model. The redshift upper limits of PG 1553+113, 3C 66A
and PKS 1424+240 are 0.78, 0.58, and 1.19 respectively. Near the TeV
energy the optical depth of VHE $\gamma$ photons might be
overestimated by Franceschini (2008) EBL model, or the second
emission component might be present in the VHE spectra and  lead the
intrinsic photon index harder than the Fermi ones.
\end{abstract}

%________________________________________________________________

\section{INTRODUCTION}
Blazars are a subclass of AGN characterized by strong non-thermal
radiation across the entire electromagnetic spectrum, from radio to
very high energy (VHE) bands. They include Flat-spectrum radio-loud
quasars (FSRQs) and BL Lac objects.  BL Lac objects usually present
no or weak emission lines, and their redshifts are not easily
determined. However, the redshift is crucial for the understanding
of VHE emissive properties of these sources due to EBL
absorption(\cite{hauser01}).

PG1553+113 is a high-frequency peaked BL Lacertae object (HBL)
(\cite{Giommi95,Beckmann:2002:BeppoSAXBLLacs}). 3C 66A is classified
as an intermediate-frequency peaked BL Lac (IBL) due to its
synchrotron peak being between $10^{15}$ and $10^{16}$ Hz
(\cite{Perri03}). The position of the synchrotron peak of PKS
1424+240 has not been measured, but it can be constrained from
optical and X-ray data to be between $10^{15}$ Hz and $10^{17}$ Hz.
Therefore, PKS 1424+240 is either regarded as an IBL
(\cite{Nieppola06}). Their redshifts are not determined as yet. The
redshift of PG1553+113 was 0.36 (\cite{Miller:1983:PG1553z}), and
then was wrong due to the misidentification of emission line
(\cite{Falomo:1990:PG1553}). Subsequent observations did not reveal
any spectral features (\cite{Falomo:1990:PG1553};
\cite{Falomo:1994:SpecBlazars}; \cite{Carangelo:2003:SpecBLLacs}).
Hubble Space Telescope (HST) images do not also determine its
redshift
(\cite{Sbarufatti:2005:Redshifts};\cite{Sbarufatti:2006:ESO-12Redshifts};
\cite{Treves:2007:PG1553z}), and only give a lower limit of $z>0.78$
(\cite{Sbarufatti:2005:Redshifts}). Sbarufatti et al. (2006) used
the spectra of ESO VLT to give a limit of $z>0.09$. \citet{Mazin07}
also gave an upper limit on the redshift of PG 1553+113 as z $<$
0.69. For 3C 66A, the redshift was estimated to be 0.444
(\cite{Miller78}; \cite{Lanzetta93}), and exits large uncertain
(\cite{Bramel05}). Recently, \citet{Finke08} give a lower limit of
$z>0.096$. For PKS 1424+240, \citet{Scarpa95} derived a limit of z
$>$ 0.06, and Sbarufatti et al. (2005) gave a limit of z $>$ 0.67.
Recently, \citet{Acciari10} deduced the redshift to be less than
0.66.

The EBL consists of the sum
of the starlight emitted by galaxies through the history of the
Universe, and includes an important contribution from the first
stars (\cite{hauser01}). The VHE photons will be absorbed by the EBL through pair production.
We can use VHE blazars with redshifts to study the EBL density. On the contrary, we can use the EBL model to limit
the redshifts and intrinsic spectra of VHE blazars.

To date, 35 AGN sources have been detected at TeV
energies\footnote[1]{update see:
http://www.mppmu.mpg.de/\~{}rwagner/sources/}. The observed spectra
have power law shapes with the index $\Gamma \geq 2$ , in which
distant sources have large $\Gamma$, up to 4 (e.g.
\cite{acciari09_0806}; \cite{Albert07b}; \cite{albert08_279};
\cite{aharonian05}). The Fermi Gamma-ray Space Telescope has
detected their emissive spectra between 20MeV and 300GeV. Assuming a
single spectral index, \citet{Abdo09_TeV} extrapolate the Fermi
spectrum up to 10 TeV as an intrinsic VHE spectrum, and find that
the break of the observed TeV spectra is consistent with the
absorption predicted by the minimal EBL density model. For a TeV
source, the presence of a break between the Fermi and VHE energy
range might be caused by some internal or external factors. The
internal factors include a break of emitting particle distribution
or an intrinsic absorption caused by strong optical-infrared
radiation within the source (Donea \& Protheroe 2003). The external
factors usually refer to the cosmic attenuation effect. Furthermore,
it is difficult to well predict the intrinsic spectrum from
simultaneous multi-wavelength observations because of the complexity
of VHE emission mechanism. In this work, we assume that the Fermi
spectral index measured by Fermi-LAT is the lower limit of intrinsic
VHE spectral index for TeV blazars. In the other words, form the
Fermi energy range to the VHE energy range, the photon index can
only be softened, not hardened (of course, there are exceptions, for
example, existing a new emitting component \citep{Yang10}, or
presence of monochromatic radiation fields within the source
\citep{Aharonian08}). Then, we limit the redshifts of three blazars
based on the observed GeV and VHE spectra. \textbf{}

In \S\ 2 we describe the method to limit the redshift assuming a
specific EBL model, and then apply the method to three blazars.
Discussions and conclusions are presented in \S\ 3.

\section{THE Methods}

The VHE $\gamma$-ray absorption by the EBL is caused by the pair
production of photon-photon collision. The observed VHE $\gamma$-ray
flux is given by
\begin{eqnarray}
f_{obs}(E_{\gamma}) = e^{-\tau(E_{\gamma})}f_{int}(E_{\gamma})\ ,
\end{eqnarray}
where $\tau(E_{\gamma})$ is the optical depth, $E_{\gamma}$ is the
observed VHE $\gamma$-ray photon energy, $f_{obs}(E_{\gamma})$ and
$f_{int}(E_{\gamma})$ are observed and intrinsic flux respectively.
Therefore, we can determine the opacity of VHE gamma-rays through
the observed and intrinsic spectra. Now we derive an upper limit of
the redshift from the opacity by assuming a specific EBL model.
Recently, many EBL models are available in the literature
\citep{Totani02,kneiske04,primack05,Stecker06,franceschini08,Raue08,Gilmore09,Finke10}.
In this work, we adopt the EBL model of \citet{franceschini08}. The
model includes evolutionary effects and the lowest level of the EBL
intensity over the concerned range(0.1-10$\mu m$, see the Fig. 7. of
Finke et al. 2010). The model is consistent with the lower limits
from galaxy counts and upper limits from observations of TeV
blazars. Therefore, using the minimum EBL intensity and the hardest
intrinsic VHE spectrum, we obtain the upper limits of three objects'
redshifts.

If the observed Fermi spectral index represents the hardest limits
of intrinsic VHE ones, the EBL will cause the difference of the
observed Fermi and VHE spectral indices. The difference, $\Delta
\Gamma= \Gamma_{VHE}- \Gamma_{Fer}$, will increase with the
redshift. For example, M 87 and Cen A with low redshifts have
$\Delta \Gamma \approx $ 0, while blazars with redshifts greater
than 0.1 show $\Delta \Gamma \geq $ 1.5(Abdo et al. 2009b).

In the VHE bands, PG 1553+113 is detected by the HESS and MAGIC
telescopes without simultaneous Fermi observation
(\cite{HESS:2006:PG1553,HESS:2008:PG1553VLT,Albert07a,Albert09}). 3C
66A is observed by the VERITAS (\cite{Acciari09}), and its neighbor
3C 66B is a possible source of VHE emission (\cite{Tavecchio08}).
Recently the MAGIC favors 3C 66B as a VHE source and excludes 3C 66A
at an 85\% confidence level (\cite{Aliu09}). However, 3C 66A has the
near-simultaneous observation of the Fermi and VERITAS in the
2008-2009 season, we favor 3C 66A as a VHE source. For the PKS
1424+240, \citet{Acciari10} reported the first detection of VHE
gamma-ray emission above 140 GeV band.

In the Fermi energy ranges, PG 1553+113 is a bright source detected
by the Fermi LAT (\cite{Abdo09}) and has a photon index of $1.70\pm
0.06$ (Since the VHE and Fermi observations are not simultaneous for
PG 1553+113, we will discuss this issue in the Section 3.). 3C 66A
and PKS 1424+240 are also detected by the Fermi survey in the first
three months and have an index of $1.97\pm 0.04$ and $1.80\pm 0.07$,
respectively (\cite{Abdo09}). 3C 66A has the near-simultaneous
observation of the Fermi and VERITAS. For PKS 1424+240,
\citet{Acciari10} discovered its VHE emission and carried out
simultaneous multi-wavelength observations, in which Fermi
observations obtain the photon index of $1.73\pm 0.07$. In this
work, we adopt the near-simultaneous or simultaneous observations to
discuss the upper limits of redshift for 3C 66A and PKS 1424+240.

We derive the variation of the optical depth with the redshift for
different VHE bands using the \citet{franceschini08} EBL models, in
which the method of linear interpolation is used. PG 1553+113 has 22
observed VHE bands, while 3C 66A and PKS 1424+240 have less observed
VHE bands.

\begin{figure}
   \centering
   \includegraphics[width=10cm]{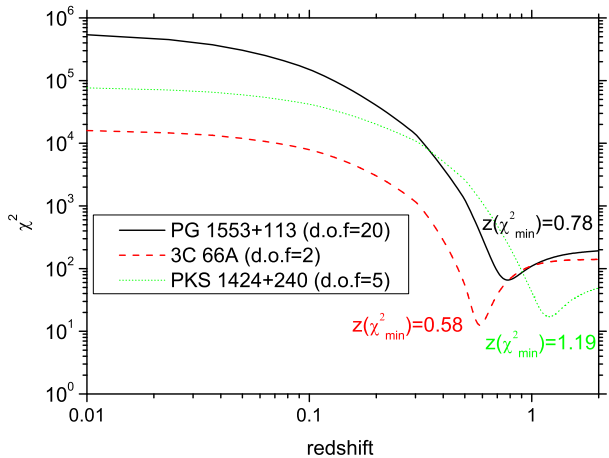}
      \caption{Change of $\chi^2$ with reshift for PG 1553+113, 3C 66A, and PKS 1424+240. The
      solid line indicates PG 1553+113, the dashed line
      denotes 3C 66A, and the doted line is PKS 1424+240. We also label the redshifts at the minimum $\chi^2$.
              }
         \label{fig.1}
   \end{figure}

Using the optical depth for specific observed VHE bands, we can
correct the intrinsic spectra to get the expected spectra. We then
calculate the $\chi^2$ between the expected and observed spectra for
different redshifts shown in the Fig.\ref{fig.1}. Finding the
redshift with the minimum $\chi^2$, we get z =0.78 for PG 1553+113,
z = 0.58 for 3C 66A, and z=1.19 for PKS 1424+240. We assume there is
no cutoff across the intrinsic GeV and VHE bands, the extrapolated
VHE spectra represent the upper limits. Therefore, our obtained
redshifts should be the upper limits.
%\clearpage
%                                              One column figure
%----------------------------------------------------------- S_vib
%
%______________________________________________________________

\section{DISCUSSIONS AND CONCLUSIONS}

\begin{figure}
   \centering
   \includegraphics[width=10cm]{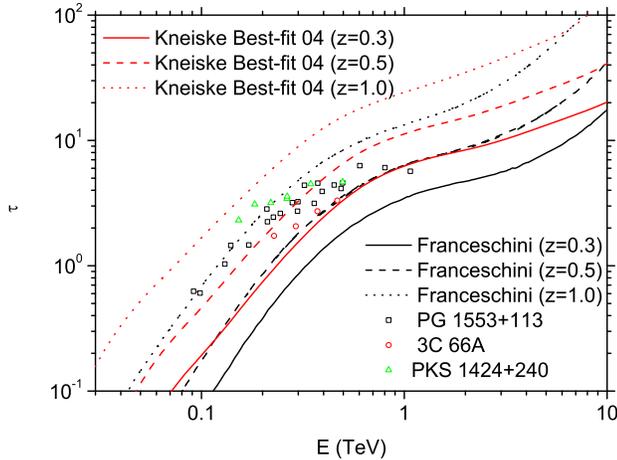}
      \caption{Relation of derived $\tau$ with VHE bands.
The open squares indicate PG 1553+113, the black circles represent
3C 66A, and the triangles are PKS 1424+240. The lines show the
results of the EBL models respectively given by
\citet{franceschini08}(black) and \citet{kneiske04} (red) under the
redshifts of 0.3, 0.5 and 1.0.
              }
         \label{fig.2}
   \end{figure}

In the Fig.\ref{fig.2}, we plot the optical depths $\tau$ of three
Blazars at different VHE bands $E_{\gamma}$. The model optical
depths given by \citet{franceschini08} (black line, almost lowest
EBL intensity over the concerned range) and Best-fit
\citet{kneiske04} (red line, moderate level) are also shown in this
figure. Since the intensity of \citet{franceschini08} EBL model is
lower than one of the \citet{kneiske04} at the range of
0.1-10$\mu$m, which have larger contribution to the absorption of
VHE photons (interaction cross section of pair production sharply
peak at $\lambda(\mu m) \approx 1.24 E_{\gamma} (TeV)
$\citep{Guy00,hauser01} for the EBL photons). For the same VHE
photons the \citet{franceschini08} EBL model give smaller optical
depth than the \citet{kneiske04} model. Therefore, using the
\citet{franceschini08} EBL model the deprived upper limits of
redshifts will be larger for the same optical depths. We find that
the derived $\tau$ have flat trend with $E_{\gamma}$ than ones given
by these models. The derived $\tau$ at lower energy bands are close
to the model values with larger redshift. It imply that the optical
depth at higher energy $E_{\gamma}$ might be overestimated by their
EBL models. The EBL at optical-infrared wavelengths is more
transparent for TeV sources than one predicted previously. Of
course, we can not rule out a flux break at TeV band caused by the
second emission component. If the intrinsic spectra have a hard
change at TeV band, the optical depths will be underestimated.

The redshifts of three blazars derived by the minimum $\chi^2$ are
compatible with the \citet{franceschini08} EBL model ones shown in
the Fig.\ref{fig.2}. For PG 1553+113, our derived redshift is
consistent with the lower limit given by
\citet{Sbarufatti:2005:Redshifts} under the HST snapshot survey.
\citet{Abdo09:PG1553} use the same method to derive the redshift
with $z < 0.75$. We note the different redshift to be caused by  the
selected VHE data from Aharonian et al. (2006). \citet{Mazin07} also
made an upper limit on the redshift of PG 1553+113 as z $<$ 0.69.
They used an argument that the VHE intrinsic photon index cannot be
harder than $\Gamma=$ 1.5 instead of the GeV ones observed by the
Fermi. The different results mainly come from different EBL model
used. For the EBL photons of 0.1-10$\mu$m  which cause serious
absorption of the VHE photons, their intensity given by
\citet{franceschini08} model is lower than one used by
\citet{kneiske04} model. For PG 1553+113, we use 22 VHE data to
derive the redshift, the result might be reliable. However, for 3C
66A and PKS 1424+240, the derived redshift has large uncertainty due
to less VHE data. \citet{Acciari10} deduced that the redshift of PKS
1424+240 is less than 0.66. The difference with our result comes
from the adopted EBL models.

Because Blazars are variable sources, the simultaneous observations
at GeV and VHE range are reliable to constrain the redshifts of BL
Lacs using this method. For the 3C 66A, there is near-simultaneous
Fermi LAT and VERITAS observations. VERITAS observed 3C 66A for 14
hr from September 2007 through January 2008 and 46 hr from September
through November 2008. Note that there were 1431 excess events
detected during the period from MJD 54740 (October 1 2008) through
MJD 54749 (October 10 2008), which accounts for 80\% of the
total\citep{Acciari09}. Thus, while the spectrum adopted here is for
the full data set, it is dominated by this period. Fermi spectrum
adopted by us is from August 4 to October 30 2008\citep{Abdo09}. We
assume that the VERITAS and Fermi observations are in the same
state. For the PKS 1424+240, \citet{Acciari10} also provided the VHE
and contemporaneous Fermi LAT observations. The VHE flux is steady
over the observation period from February 19 2009 to June 21, the
LAT data overlapping with the VERITAS observations were analyzed by
them \citep{Acciari10}. PG 1553+113 has four VHE data
sets\citep{HESS:2006:PG1553,HESS:2008:PG1553VLT,Albert07a,Albert09},
but they are not simultaneous with the Fermi LAT ones. The VHE
photon index of PG 1553+113, respectively reported by HESS and MAGIC
with $\Gamma_{obs}=4.0 \pm 0.6$, $\Gamma_{obs}=4.5 \pm 0.3$,
$\Gamma_{obs}=4.2 \pm 0.3$, and $\Gamma_{obs}=4.1 \pm 0.3$, are very
similar. The deviations of energy spectral index from the average
value are less than 15\%. Normalizing the flux at 300 GeV, we find
that the change of flux at 300 GeV are not more than 30\%.
Uncertainty due to non-simultaneous observation is discussed. We
calculate the relative error of upper limits of redshift caused by
variability of energy index and flux at VHE range shown in the Table
1. For blazars the variability of index and flux should be
simultaneous, separately discussing the errors caused by variability
of index and flux in this work might be a simplified one. PG
1553+113 has 22 data points observed by HESS and MAGIC, we use a
power-law spectrum to fit these points with the energy index
$\Gamma_0=2.20$ and the flux $F_0= 3.54 \times 10^{-11} cm^{-2}
s^{-1} TeV^{-1}$ at 300 GeV. Based on $\Gamma_0$, $F_0$ and Fermi
data, we calculate the upper limit $Z_0$ of redshift. We define
$V_F=F/F_0$ and $V_I=\Gamma/\Gamma_0$ to show the variability of the
energy index and flux. We fix the flux $F_0$ at 300 GeV and change
the energy index shown by $V_I$, , then we obtain the upper limits
$Z$ and the relative errors $RE=|Z-Z_0|/Z_0$. We find that $RE$ is
less than 2\% when $\Gamma$ changes about 15\%. Similarly we fix the
energy index and change the flux described by $V_F$ to calculate $Z$
and $RE$. We find that $RE$ is less than 12\% when the flux at 300
GeV changes 30\%.

\begin{table*}
\caption{Relative error of upper limits of redshift if the flux or
energy index vary at VHE range for PG 1553+113 } \centering
\begin{tabular}{@{}lccccccc}
 \hline
  $V_{I}$                 & 0.85 & 0.90 & 0.95 & 1.00  & 1.05 & 1.10 & 1.15\\
   \hline
Z                        & 0.76 & 0.76 & 0.76  & 0.76  & 0.74  & 0.74  & 0.74\\
RE                        & 0\% & 0\% & 0\% & 0\%  & 2\% & 2\% & 2\%\\
 \hline
 $V_{F}$                  &  0.7  &  0.8   & 0.9  &  1.0  &  1.1  &  1.2  &  1.3\\
   \hline
Z                       & 0.85 & 0.81 & 0.78 & 0.76  & 0.73 & 0.70 & 0.69\\
RE                        & 12\% & 7\% & 4\% & 0\%  & 4\% & 7\% & 9\%\\
   \hline

\label{table} \\
\end{tabular}\\
\end{table*}

With our obtained upper limits of redshifts, the de-absorbed
intrinsic VHE spectra are hard. If the redshifs are real, the
cascade emission will significantly contribute to the GeV spectrum.
The GeV radiation is produced by $e^{\pm}$ pairs scattering the
cosmic microwave background photons on the way from the source to
us, where the $e^{\pm}$ are created by VHE photons interacting with
the EBL. However, the GeV emission greatly depends on the EBL model
and intergalactic magnetic field which has large uncertainty
\citep{Dai02,Fan04,Yang08}. Detailed discussions are complicated and
need further observations in the future.

%----------------------------------------------------------- S_vib
%
%______________________________________________________________

We thank the referee for a very helpful and constructive report that
helped to improve our manuscript substantially. We acknowledge the
financial supports from the National Natural Science Foundation of
China 10673028 and 10778702, the National Basic Research Program
of China (973 Program 2009CB824800) and the
Policy Research Program of Chinese Academy of Sciences (KJCX2-YW-T24).

\end{document}